\documentclass[12pt]{article}
\usepackage{latexsym}
\usepackage{amsmath}
\usepackage{amssymb}
\catcode`\@=11
\newcount\hour
\newcount\minute
\newtoks\amorpm \hour=\time\divide\hour by 60\minute
=\time{\multiply\hour by 60 \global\advance\minute by-\hour}
\edef\standardtime{{\ifnum\hour<12 \global\amorpm={am}%
        \else\global\amorpm={pm}\advance\hour by-12 \fi
        \ifnum\hour=0 \hour=12 \fi
        \number\hour:\ifnum\minute<10
        0\fi\number\minute\the\amorpm}}
\edef\militarytime{\number\hour:\ifnum\minute<10
0\fi\number\minute}
\def\draftlabel#1{{\@bsphack\if@filesw {\let\thepage\relax
   \xdef\@gtempa{\write\@auxout{\string
      \newlabel{#1}{{\@currentlabel}{\thepage}}}}}\@gtempa
   \if@nobreak \ifvmode\nobreak\fi\fi\fi\@esphack}
        \gdef\@eqnlabel{#1}}
\def\@eqnlabel{}
\def\@vacuum{}
\def\marginnote#1{}
\def\draftmarginnote#1{\marginpar{\raggedright\scriptsize\tt#1}}
\def\draft{
        \pagestyle{plain}
        \overfullrule=2pt
        \oddsidemargin -.1truein
        \def\@oddhead{\sl \phantom{\today\quad\militarytime} \hfil
        \smash{\Large\sl DRAFT} \hfil \today\quad\militarytime}
        \let\@evenhead\@oddhead
        \let\label=\draftlabel
        \let\marginnote=\draftmarginnote
        \def\ps@empty{\let\@mkboth\@gobbletwo
        \def\@oddfoot{\hfil \smash{\Large\sl DRAFT} \hfil}
        \let\@evenfoot\@oddhead}
        \def\@eqnnum{(\theequation)\rlap{\kern\marginparsep\tt\@eqnlabel}%
        \global\let\@eqnlabel\@vacuum}  }

\renewcommand{\theequation}{\thesection.\arabic{equation}}
\renewcommand{\thefootnote}{\fnsymbol{footnote}}
\newcommand{\newsection}{    
\setcounter{equation}{0}\section}
\def\appendix#1{\addtocounter{section}{1}\setcounter{equation}{0}
\renewcommand{\thesection}{\Alph{section}}
\section*{Appendix \thesection\protect\indent \parbox[t]{11.15cm}{#1}}
\addcontentsline{toc}{section}{Appendix \thesection\ \ \ #1}}

\def \la {\label}

\def\be{\begin{equation}}
\def\ee{\end{equation}}

\def\nat {{\natural}}

\newcommand{\cont}[1]{{}_{#1}{}^{#1}}

\catcode`\@=11

\def\bea{\begin{eqnarray}}
\def\eea{\end{eqnarray}}
\def\beann{\begin{eqnarray*}}
\def\eeann{\end{eqnarray*}}
\def\beq{\begin{equation}}
\def\eeq{\end{equation}}
\def\ba{\begin{array}}
\def\ea{\end{array}}
\def\ben{\begin{enumerate}}
\def\een{\end{enumerate}}
 
 \def\m {\mu}

 \def \la {\label}
 \def\be{\begin{equation}}
\def\ee{\end{equation}}

\def \la {\label}

\font\mybb=msbm10 at 11pt

\def\bb#1{\hbox{\mybb#1}}

\def\bR {\bb{R}}

\def\bC {\bb{C}}

\def\e  {\epsilon}

\def \ee {\epsilon}

\def \m {\mu}

\def\lc{\lrcorner}

\def\be{\begin{equation}}
\def\ee{\end{equation}}

\def \la{\label}
\newcommand{\al}{\alpha}
\newcommand{\ie}{{\em i.e.}~}
\newcommand{\eg}{{\em e.g.}~}

\begin{document}
\date{November 2002}
\begin{titlepage}
\begin{center}
\vspace*{-1.0cm}
\hfill hep-th/0612148 \\
\hfill KUL-TF-06/33 \\
\hfill UB-ECM-PF-06-43 \\

\vspace{2.0cm} {\Large \bf Aspects of spinorial geometry} \\[.2cm]

\vspace{1.5cm}
 {\large  U. Gran$^1$, J. Gutowski$^2$,  G. Papadopoulos$^3$ and D. Roest$^4$}

\vspace{0.5cm}

${}^1$ Institute for Theoretical Physics, K.U. Leuven\\
Celestijnenlaan 200D\\
B-3001 Leuven, Belgium\\

\vspace{0.5cm}
${}^2$ DAMTP, Centre for Mathematical Sciences\\
University of Cambridge\\
Wilberforce Road, Cambridge, CB3 0WA, UK

\vspace{0.5cm}
${}^3$ Department of Mathematics\\
King's College London\\
Strand\\
London WC2R 2LS, UK\\

\vspace{0.5cm}
${}^4$ Departament Estructura i Constituents de la Materia \\
    Facultat de F\'{i}sica, Universitat de Barcelona \\
    Diagonal, 647, 08028 Barcelona, Spain \\

\end{center}

\vskip 1.5 cm
\begin{abstract}

We review some aspects of the spinorial geometry approach to the classification of supersymmetric solutions
of supergravity theories. In particular, we explain how spinorial geometry can be used to
express the Killing spinor equations in terms of a linear system for the fluxes and the geometry of spacetime.
The solutions of this linear system express some of the fluxes in terms of the spacetime geometry and
determine the conditions on the spacetime geometry imposed by supersymmetry.
We also present some of the recent applications like the classification of maximally supersymmetric $G$-backgrounds in IIB,
this includes the most
general pp-wave solution preserving 1/2 supersymmetry,
and the classification of $N=31$ backgrounds in ten and eleven dimensions.

\end{abstract}

\end{titlepage}
\newpage
\setcounter{page}{1}
\renewcommand{\thefootnote}{\arabic{footnote}}
\setcounter{footnote}{0}

\setcounter{section}{0}
\setcounter{subsection}{0}
\newsection{Introduction}

Ten- and eleven-dimensional supergravities are the low energy effective theories of string and M-theory. As such they
have proved instrumental for exploring the brane solitons of string/M-theory, string dualities, string/M-theory
compactifications and more recently the AdS/CFT correspondence, see \eg{}\cite{townsend,strominger,townsendhull,maldacena}.
All these applications have been mediated by the use of a special class
of supergravity solutions, those solutions that admit Killing spinors, and thus preserve some of the spacetime supersymmetry.
Most of these solutions have been constructed using Ans\"atze adapted to the requirements of  physical problems.
However, it has become increasingly clear that it will be advantageous to classify all supersymmetric supergravity solutions.

Our knowledge of the space of all supersymmetric supergravity  solutions in ten and eleven dimensions
 is rather limited. It was a surprise for example that IIB supergravity admits an additional maximally supersymmetric solution \cite{bfhp,bfhp2}, which has found
 applications in the AdS/CFT correspondence \cite{bmn}. The maximally supersymmetric solutions of $D=10$ and $D=11$ supergravities
 have only recently been classified \cite{jfgp,jfgp2}. The Killing spinor equations (KSE) for\footnote{$N$ denotes the number of Killing spinors a background admits.}
$N=1$ backgrounds of $D=11$ supergravity have also been solved in \cite{gjs,gjs2} using $G$-structures.
It is expected that there are many more supersymmetric solutions in ten and eleven dimensions that remain to be uncovered. In lower
dimensional supergravities much more progress has been made initiated by Tod in \cite{tod}, see \eg{}\cite{hullpakis}.

The classification of supersymmetric solutions is an attractive geometric problem.
The  Riemannian analogue is
to find the manifolds that admit parallel spinors. These
are identified as a consequence of
Berger classification of irreducible Riemannian manifolds. In supergravity, the Berger classification cannot be applied
because of the presence of fluxes.

Spinorial geometry, proposed in \cite{ggp}, is a direct and effective method of solving the KSE of supergravity theories.
These consist of a parallel transport equation for the supercovariant connection and algebraic conditions
derived  from the vanishing condition of the supersymmetry transformations of the fermions.
The spinorial geometry method is based on the following ingredients:

\begin{itemize}

\item The gauge symmetry of the Killing spinor equations.

\item A description of spinors in terms of forms.

\item An oscillator basis in the space of spinors.

\end{itemize}

The gauge transformations are those that leave the form of the KSE invariant. The gauge group of the KSE is typically a $Spin$ group.
The holonomy of the supercovariant connections \cite{hull,duff,gpdt,gpdt2} on the other hand, is typically an $SL$ group, which contains the gauge group, and does not preserve the form of the KSE.
Backgrounds related by gauge transformations are identified. The gauge transformations can be used
to orient the Killing spinors along some directions. This has proved instrumental for solving the KSE
with small and near maximal number of supersymmetries.

An elegant way to represent spinors is in terms of (multi)-forms proposed by Cartan. This notation gives a geometric insight into the
KSE and simplifies many of the computations.

The oscillator basis in the space of spinors is used to write the KSE, or their integrability
conditions, as a linear system for the fluxes and the geometry of the spacetime. The linear system can then be solved to express
some of the fluxes in terms of the geometry and to determine the conditions on the spacetime geometry imposed
by supersymmetry \cite{syst11,systiib}.

Effective use of the spinorial geometry method requires all three of the above ingredients. The method has, for example, been used to solve
the KSE of $N=1$ IIB backgrounds in \cite{ugjggp,ugjggp2}, and to considerably simplify the analogous  computation for $N=1$ backgrounds in eleven-dimensional supergravity \cite{ggp}, originally done in \cite{gjs,gjs2}.
It has also been applied to classify, under some mild assumptions, the geometry of all supersymmetric heterotic string \cite{ugplgp}
  and $N=2$ common sector backgrounds \cite{ugplgp2}. More recently, the spinorial geometry method has been adapted to near maximally supersymmetric backgrounds and it has been
used to show that $N=31$ IIB and $N=31$ (simply connected) eleven-dimensional backgrounds are maximally supersymmetric\footnote{It would be of interest to re-derive these results using $G$-structures.} \cite{iibpreons,11preons}.
Some other applications can be found in \cite{kappa,giib}. In particular, in \cite{giib} all the supersymmetric IIB backgrounds
which admit the maximal number of Killing spinors invariant under some non-trivial Lie subgroup of $Spin(9,1)$ have been classified.

In this review, we first describe how spinorial geometry can be used to write the KSE for any number of Killing spinors
in terms of a linear
system for the fluxes and the geometry of spacetime, \ie{}we give a systematic way to
solve the KSE for any background. Then
we present two applications. The first application is the classification of
maximally supersymmetric IIB $G$-backgrounds.
Some of these can be thought of as the vacua of IIB string compactifications.  These also include the most
general pp-wave solution preserving (at least) 16 supersymmetries. The other application is the classification
of $N=31$ IIB and $D=11$ supergravity backgrounds. In particular, we show that in both cases the $N=31$ backgrounds
 admit an additional Killing spinor and
therefore they are (locally) isometric
to maximally supersymmetric ones.

This review is organized as follows: In section two, we present the systematics of spinorial geometry. In section three,
we give the classification of maximally supersymmetric IIB $G$-backgrounds. In section four, we investigate the backgrounds with
31 supersymmetries, and in section five, we give our conclusions.

\newsection{The linear system of Killing spinor equations}

To explain the construction of the linear system associated to the KSE, we shall first
give a description of spinors in terms of forms. Then we shall use this to construct the linear system.
Similarly,  we shall also  discuss how the integrability conditions of the KSE
reduce to a linear system for the bosonic supergravity field equations.

\subsection{Expansion in terms of basis of forms}

A description of spinors in terms of forms \cite{lawson,harvey,wang,ggp} is advantageous because it gives a geometric insight into the nature
of spinors. In turn  this makes many computations necessary for the classification of supersymmetric backgrounds straightforward,
\eg the construction of the linear systems that we explain below and the computation
of spinor isotropy groups.

As a paradigm, let us construct the Majorana representation of $Spin(10,1)$. For this we begin with the spinor representation of $Spin(10)$.
Let $U=\bC<e_1,\dots,e_5>$ be a complex vector space spanned by the orthonormal vectors $e_1,\dots, e_5$ and equipped with the Hermitian
inner product
\be
<z^i e_i, w^j e_j>=\sum_{i=1}^{10} \bar z^i w^i~,
\ee
where $\bar z^i$ is the standard complex conjugate of $z^i$ in $U$.
The space of Dirac $Spin(10)$ spinors is $\Delta_c=\Lambda^*(U)$.
The above inner product can be easily extended to $\Delta_c$ and it is called
the Dirac inner product on the space of spinors.
The gamma matrices  act on $\Delta_c$ as
\bea
\Gamma_i\eta= e_i\wedge\eta+e_i\lc \eta~,~~ i\leq 5 \,, \qquad
\Gamma_{5+i}\eta=ie_i\wedge\eta-ie_i\lc \eta~,~~ i\leq 5~,
\eea
where $e_i\lc$ is the adjoint of $e_i\wedge$ with respect to $<,>$.
The linear maps $\Gamma_i$ are Hermitian with respect
to the inner product $<,>$, $<\Gamma_i \eta, \theta>=<\eta, \Gamma_i\theta>$,
and satisfy the Clifford algebra relations
$\Gamma_i\Gamma_j+\Gamma_j \Gamma_i=2 \delta_{ij}$. It will be convenient to
define a Hermitian basis for the $\Gamma_i$ as
 \be
 \Gamma_{\bar\al}={1\over \sqrt{2}} (\Gamma_\al+i \Gamma_{\al+5})~,~~~~\al=1,\dots,5~,
 \ee
and $\Gamma^\al=g^{\al\bar\beta} \Gamma_\beta$, where $g_{\al\bar\beta}=\delta_{\al\bar\beta}$. The Dirac $Spin(10,1)$ representation,
$\Delta_c=\Lambda^*(U)$, is constructed by
identifying $\Gamma_0 = \Gamma_1\dots \Gamma_{\nat}$, where $\Gamma_\nat=\Gamma_{10}$.

The Majorana spinor inner product on $\Delta_c$ is
\be
B(\eta, \theta)=<B(\bar\eta), \theta>~,
\ee
where $\bar\eta$ is the standard complex conjugate of $\eta$ in
$\Lambda^*(U)$ and $B=\Gamma_6\dots\Gamma_{\nat}$.
It is easy to verify that $B(\eta, \theta)=-B(\theta, \eta)$, \ie{}$B$ is
skew-symmetric.  The Majorana condition can be easily imposed by setting
\be
\bar\eta= \Gamma_0 B(\eta)~,~~~~~~\eta\in \Delta_c~.
\la{maj}
\ee
The Majorana spinors $\Delta_{32}$ of eleven-dimensional supergravity are those spinors in $\Delta_c$
which obey the Majorana condition (\ref{maj}).
The $Pin(10)$-invariant inner product $B$ induces a $Spin(10,1)$ invariant inner product
on $\Delta_{32}$ which is the usual skew-symmetric inner product on the space of spinors
of eleven-dimensional supergravity.

A consequence of the construction above is that any Majorana spinor of $Spin(10,1)$ can be written as
\bea
\epsilon &=& f (1+e_{12345})+ i
g (1- e_{12345}) +\sqrt{2} u^i (e_i+
\tfrac{1}{4!}\epsilon_i{}^{jklm} e_{jklm})\nonumber \\&& + i \sqrt{2} v^i
(e_i-\tfrac{1}{4!}\epsilon_i{}^{jklm} e_{jklm})  +
\tfrac{1}{2} w^{ij} (e_{ij}-\tfrac{1}{3!} \epsilon_{ij}{}^{klm}
e_{klm}) \nonumber \\&&+ \tfrac{i}{2} z^{ij}(e_{ij}+\tfrac{1}{3!}
\epsilon_{ij}{}^{klm} e_{klm}) ~, \la{genspinor}
\eea
where $i,j, \ldots = 1, \ldots, 5$ and $f,g, u^i, v^i, w^{ij}$ and $z^{ij}$ are real spacetime
functions. Clearly, $\epsilon$ can also be expressed in the Hermitian basis
 \bea
 \epsilon &=& f^I \sigma_I \,, \qquad \sigma_I = (1, e_i, e_{ij}, e_{ijk},
 e_{ijkl}, e_{12345} ) \,,
 \label{hermbasis}
 \eea
where $f^I$ are complex functions.
The six types of spinors $\sigma_I$ correspond
to the irreducible representations of $U(5)$ on $\Lambda^*(\bC^5)$. This basis of spinors, for reasons that we shall not explain here,
is referred to as a timelike basis \cite{syst11}.

Backgrounds that are related by a gauge transformation of the KSE are identified.
Because of this, if two sets of Killing spinors are related by such a gauge transformation, then they give
  rise to the
same supersymmetric background. So to classify the different supersymmetric backgrounds, one has to identify the inequivalent classes
of Killing spinors up to gauge transformations. In turn, this leads to choosing representatives of the orbits
of the gauge group in the space of spinors. Considerable simplification can be made in the various computations, if one
chooses carefully such representatives. For example, there are two types of orbits of $Spin(10,1)$ in $\Delta_{32}$,
one has stability subgroup $SU(5)$ while the other has stability subgroup
$(Spin(7) \ltimes \mathbb{R}^8) \times
\mathbb{R}$ \cite{bryant,jose}. So there are two different types of geometries that can occur in $N=1$ eleven-dimensional backgrounds.
 A representative of the $SU(5)$ orbit is
  \bea
  \epsilon&=& f (1+e_{12345}) \,.
  \label{sss}
  \eea
Compared to the general spinor in (\ref{genspinor}), this representative is much simpler. In turn, the linear system
associated with (\ref{sss}) is rather
simple and can be straightforwardly  solved \cite{ggp}. The use of the gauge group
is essential for the analysis of solutions preserving a large number of supersymmetries as well.

The same analysis can be done for the integrability conditions
${\cal I}\epsilon=0$ of a Killing spinor $\epsilon$. Since these
conditions are linear, we have
\bea
{\cal I} \epsilon &=& f^I {\cal I} \sigma_I \,.
\eea
Therefore to find which field equations are
determined by the KSE, it suffices to compute ${\cal I} \sigma_I$.
In this way the integrability conditions give rise to a linear system in terms
of the field equations. The solution to this linear system shows which field
equations are already satisfied or related to others, and which field
equations are independent and still have to be imposed. See \cite{syst11} for
the explicit expressions for each $e_{i_1 \cdots i_I}$ and applications of
this linear system in some $N=1,2$ and $4$ examples.

The Majorana representation of $Spin(10,1)$ can also be constructed from the spinor representations
of $Spin(9,1)$. This leads to another (null) basis  in the space of $Spin(10,1)$ spinors from the timelike basis constructed above.
One advantage of the null basis is that one can easily investigate the cases where  spinors are in the
 $(Spin(7) \ltimes \mathbb{R}^8) \times
\mathbb{R}$ orbit. The null basis is also the preferred basis to investigate the KSE of IIB
and type I supergravities. Details of the construction of this basis and some applications
can be found in \cite{ugjggp,systiib}.

\subsection{Systematics of Killing spinor equations}

To explain the construction of the linear system associated with the KSE,
we observe, using (\ref{hermbasis}), that
\bea
{\cal D}_A\epsilon= \partial_A f^I \sigma_I + f^I {\cal D}_A \sigma_I \,,
\la{genspcov}
\eea
where ${\cal D}$ is the supercovariant connection of eleven-dimensional supergravity.
Thus the KSE reduce to the
evaluation of ${\cal D}$ on the basis spinors
$\sigma_I$. So it remains to compute these and express the result in the basis
(\ref{hermbasis}). To do this, first write
 \be
  \sigma_{i_1\cdots i_I}=e_{i_1 \cdots i_I} = \frac{1}{2^{I / 2}} \Gamma^{\bar i_1}
  \cdots \Gamma^{\bar i_I} 1 \,,
 \label{basis-element}
 \ee
where the indices $i_1, \ldots, i_I$ pick out $I$ holomorphic
indices (with $0 \leq I \leq 5$) from the range $\al = 1, \ldots, 5$.
It will be convenient to distinguish between the indices that do
appear in the basis element (\ref{basis-element}) and those that do
not: we split the holomorphic indices $\al$ into the
indices\footnote{The $i_1,\ldots,i_I$ should not be thought of as
indices in this context, but rather as fixed labels for a particular
spinor.} $a = (i_1, \ldots, i_I)$ and the remaining $5-I$ indices
$p$, and similarly for the anti-holomorphic indices $\bar \al$. Note
that $\Gamma^{\bar a}$ and $\Gamma^p$ annihilate the spinor $e_{i_1
\cdots i_I}$ while $\Gamma^{a}$ and $\Gamma^{\bar p}$ act as
creation operators. For this reason it is useful to define the new
indices $\rho, \sigma, \tau$ consisting of the combination
 \be
  \rho = (\bar a_1, \ldots, \bar a_I, p_1, \ldots,
  p_{5-I}) \,, \qquad  \bar{\rho} = (a_1, \ldots, a_I, \bar p_1, \ldots, \bar p_{5-I})
 \,,
 \ee
where $\Gamma^\rho$ and $\Gamma^{\bar \rho}$ are the annihilation
and creation operators, respectively, for the spinor  $e_{i_1 \cdots
i_I}$. Note that the indices $\al$ and $\rho$ are
 identical for $I=0$,
\ie{}for the spinor $1$. For $I > 0$, \ie{}for any other basis
element, these indices differ.

In terms of the basis\footnote{Note that in this basis $e_{i_1
\cdots i_I}$ is the Clifford algebra vacuum.}
 \be
  \{ e_{i_1 \cdots i_I}, \Gamma^{\bar \sigma_1} e_{i_1 \cdots i_I}, \ldots,
  \Gamma^{\bar \sigma_1 \cdots \bar \sigma_5} e_{i_1 \cdots i_I}
  \} \,, \label{element-basis}
 \ee
the supercovariant derivative with $A =0$ can be expanded in the
following contributions:
 \bea
 {{\cal D}}_0 e_{i_1 \cdots i_I} &=& [\tfrac{1}{2}\Omega_{0, \tau}{}^{ \tau}
 + (-1)^{I+1} \tfrac{i}{24} F_{\tau_1}{}^{ \tau_1}{}_{ \tau_2} {}^{
 \tau_2}] e_{i_1 \cdots i_I}
 + [(-1)^I \tfrac{i}{2} \Omega_{0,0 \bar \sigma}
  + \tfrac{1}{6} G_{\bar \sigma \tau}{}^{\tau}] \Gamma^{\bar \sigma} e_{i_1 \cdots i_I}
   \nonumber \\
  &+& [\tfrac{1}{4} \Omega_{0,\bar \sigma_1 \bar \sigma_2}
  + (-1)^{I+1} \tfrac{i}{24} F_{\bar \sigma_1 \bar \sigma_2 \tau}{}^{\tau}]
  \Gamma^{\bar \sigma_1 \bar \sigma_2} e_{i_1 \cdots i_I}
  + [\tfrac{1}{36} G_{\bar \sigma_1 \bar \sigma_2 \bar \sigma_3}]
\Gamma^{\bar \sigma_1 \bar \sigma_2 \bar \sigma_3} e_{i_1 \cdots
i_I} \nonumber \\ &+& [(-1)^{I+1} \tfrac{i}{288} F_{\bar \sigma_1
\cdots \bar \sigma_4}] \Gamma^{\bar \sigma_1 \cdots \bar \sigma_4}
e_{i_1 \cdots i_I}~. \la{genzcom}
 \eea
Observe that the component $\Gamma^{\bar \sigma_1 \cdots \bar
\sigma_5} e_{i_1 \cdots i_I}$ vanishes. Similarly, the expression
for $A = \rho$ read
 \bea
 {{\cal D}}_{\rho} e_{i_1
\cdots i_I}  &=& [\tfrac{1}{2}\Omega_{\rho, \sigma}{}^{ \sigma}
 + (-1)^{I} \tfrac{i}{4} G_{\rho \sigma} {}^{\sigma}] e_{i_1 \cdots i_I}
 \nonumber \\&+& [(-1)^I \tfrac{i}{2} \Omega_{\rho,0 \bar \sigma}
  + \tfrac{1}{4} F_{\rho \bar \sigma \tau}{}^{\tau}
  - \tfrac{1}{24} g_{\rho \bar \sigma} F \cont{\tau_1} \cont{\tau_2}]
  \Gamma^{\bar \sigma} e_{i_1 \cdots i_I}
   \nonumber \\
  &+& [\tfrac{1}{4} \Omega_{\rho,\bar \sigma_1 \bar \sigma_2}
  + (-1)^I \tfrac{i}{8} G_{\rho \bar \sigma_1 \bar \sigma_2} + [
    (-1)^{I+1} \tfrac{i}{12} g_{\rho [ \bar \sigma_1} G_{\bar \sigma_2 ]}] \cont{\tau}]
    \Gamma^{\bar \sigma_1 \bar \sigma_2} e_{i_1 \cdots i_I}
     \nonumber \\
&+& [\tfrac{1}{24} F_{\rho \bar \sigma_1 \bar \sigma_2 \bar
\sigma_3} -
  \tfrac{1}{24} g_{\rho [ \bar \sigma_1} F_{\bar \sigma_2 \bar \sigma_3 ]} \cont{\tau}]
  \Gamma^{\bar \sigma_1 \bar \sigma_2 \bar \sigma_3} e_{i_1 \cdots i_I}
  \nonumber \\
   &+& [(-1)^{I+1} \tfrac{i}{72} g_{\rho [ \bar \sigma_1}
G_{\bar \sigma_2 \bar \sigma_3 \bar \sigma_4 ]}] \Gamma^{\bar
\sigma_1 \cdots \bar \sigma_4} e_{i_1 \cdots i_I} \nonumber \\&+& [
 -\tfrac{1}{288} g_{\rho [ \bar \sigma_1} F_{\bar \sigma_2 \cdots \bar \sigma_5
 ]}]
 \Gamma^{\bar \sigma_1 \cdots \bar \sigma_5} e_{i_1 \cdots i_I}\,.
 \la{genrcom}
 \eea
Finally, for $A =\bar \rho$ we find
 \bea
 {{\cal D}}_{{\bar \rho}} e_{i_1 \cdots
i_I}&=& [\tfrac{1}{2}\Omega_{{\bar \rho},\sigma}{}^{\sigma}
 + (-1)^{I} \tfrac{i}{12} G_{{\bar
\rho}  \sigma} {}^{ \sigma}] e_{i_1 \cdots i_I} + [(-1)^I
\tfrac{i}{2} \Omega_{{\bar \rho},0 \bar \sigma}
  + \tfrac{1}{12} F_{{\bar
\rho} \bar \sigma \tau}{}^{\tau}]\Gamma^{\bar \sigma} e_{i_1 \cdots
i_I}
 \nonumber \\
 &+& [\tfrac{1}{4} \Omega_{{\bar
\rho},\bar \sigma_1 \bar \sigma_2}
  + (-1)^I \tfrac{i}{24} G_{{\bar
\rho} \bar \sigma_1 \bar \sigma_2}]  \Gamma^{\bar \sigma_1 \bar \sigma_2}
 e_{i_1 \cdots i_I}
 \nonumber \\&+& [\tfrac{1}{72} F_{{\bar
\rho} \bar \sigma_1 \bar \sigma_2 \bar \sigma_3}]
 \Gamma^{\bar \sigma_1 \bar \sigma_2 \bar \sigma_3} e_{i_1 \cdots i_I}~.
 \la{genrbcom}
 \eea
Observe that the components along $\Gamma^{\bar \sigma_1 \cdots \bar \sigma_4}
 e_{i_1 \cdots i_I}$
and $\Gamma^{\bar \sigma_1 \cdots \bar \sigma_5} e_{i_1 \cdots i_I}$
vanish.

It is convenient to convert the above expressions
 from  basis
(\ref{element-basis}) to the ''canonical'' basis \eqref{hermbasis}.
For this, we expand the products of $\Gamma^{\bar \rho}$ matrices,
which are creation operators on $e_{i_1 \cdots i_I}$, into a sum of
products of $\Gamma^{a}$ and $\Gamma^{\bar p}$ matrices, which are
annihilation and creation operators, respectively, on $1$. Then we
act on $e_{i_1 \cdots i_I}$ with the annihilation operators. In
particular, we have
 \bea
  {\cal D}_A e_{i_1 \cdots i_I}&=& \sum_k [{\cal D}_A e_{i_1 \cdots i_I}]_{\bar\rho_1\cdots \bar\rho_k}
  \Gamma^{\bar\rho_1\cdots\bar\rho_k} e_{i_1 \cdots i_I} \nonumber \\ &=& \sum_k \sum_{m+n=k}
 \frac{k!}{m! n!} [{\cal D}_{A} e_{i_1
 \cdots i_I}]_{a_1 \cdots a_m \bar p_1   \cdots \bar p_n}
 \Gamma^{a_1 \cdots a_m} \Gamma^{\bar p_1 \cdots
 \bar p_n} e_{i_1\dots i_I}
 \nonumber \\
 &=& \sum_k \sum_{m+n=k}
 \frac{k!}{m! n!}   \frac{(-1)^{[m/2]+nI}}{{2}^{{I / 2}-m} (I-m)!}
  \epsilon^{a_1 \cdots a_m}{}_{\bar a_{m+1} \cdots \bar a_I}
  \nonumber \\
  &&
  [{\cal D}_A e_{i_1 \cdots i_I}]_{a_1 \cdots a_m \bar p_1 \cdots \bar p_n}
   \Gamma^{\bar a_{m+1} \cdots \bar a_I \bar p_1 \cdots
  \bar p_n} 1\,,
  \la{genform}
 \eea
with the obvious restrictions $m \leq I$ and $n \leq 5-I$ and the
convention that $\epsilon_{\bar i_1 \cdots \bar i_I} = 1$. Using the
expressions  (\ref{genzcom}), (\ref{genrcom}) and (\ref{genrbcom})
for the components of ${\cal D}_A e_{i_1 \cdots i_I}$ in the basis
(\ref{element-basis}) which appear in square brackets in
(\ref{genform}), one can easily compute the components of ${\cal
D}_A e_{i_1 \cdots i_I}$ in the canonical basis
(\ref{hermbasis}). The explicit
expressions for the different basis elements are given in \cite{syst11}.

Substituting (\ref{genform}) into (\ref{genspcov}) and setting each component in the basis \eqref{hermbasis}  equal to zero, one
derives a linear system with variables the spin
connection $\Omega$ of the geometry and the fluxes that appear in the supercovariant
derivative, in this case the four-form $F$. In addition there are terms
consisting of the differential of the functions $f^I$ that define the
Killing spinors. Solving this linear system is equivalent to solving the KSE for any number of spinors.
In particular, one can derive all the conditions on a background imposed by supersymmetry.

\newsection{Maximally supersymmetric $G$-backgrounds}

Amongst the various supersymmetric IIB backgrounds are those for which the Killing spinors are invariant
 under the action
of some proper Lie subgroup $G$ of $Spin(9,1)$.
Let $\Delta^G$ be the space of $G$-invariant spinors.
If $\{ \eta_p : p=1, \dots , m \} $ is a maximal set of linearly independent
Majorana-Weyl spinors in $\Delta^G$, a basis for $\Delta^G$
is given by $\{ \eta_i : i=1, \dots 2m \} = \{ \eta_p, i \eta_p : p=1, \dots , m \}$,
so $\Delta^G$ is $2m$-dimensional.

We shall consider solutions for which the Killing spinors span $\Delta^G$;
such solutions are called maximally supersymmetric $G$-backgrounds.
These types of solutions were first investigated in \cite{ugjggp2},
where the KSE corresponding to
$G= Spin(7) \ltimes \bR^8$ ($\dim \Delta^G=2$), $G=SU(4) \ltimes \bR^8$
($\dim \Delta^G=4$) and
$G=G_2$ ($\dim \Delta^G=4$) were examined.
The integrability conditions of these examples were
then analyzed in \cite{systiib}. The remaining examples of
maximally-supersymmetric $G$-backgrounds with $G=Sp(2)  \ltimes \bR^8$ ($\dim \Delta^G=6$),
$G=(SU(2) \times SU(2)) \ltimes \bR^8$ ($\dim \Delta^G=8$), $G=\bR^8$ ($\dim \Delta^G=16$), $G=SU(3)$ ($\dim \Delta^G=8$),
$G=SU(2)$ ($\dim \Delta^G=16$)and $G=\{ 1 \}$ ($\dim \Delta^G=32$) were later constructed in \cite{giib}.

For maximally-supersymmetric $G$-backgrounds, the Killing spinors $\epsilon_i$ are given by
\be
\epsilon_i = \sum_{r=1}^{2m} f_{ir} \eta_r , \qquad i=1, \dots , 2m\,,
\ee
where $f_{ir}$ is a $2m \times 2m$ invertible real matrix,
whose components $f_{ij}$ in general are not constant, but depend on the spacetime
co-ordinates. Using
these properties of $f$, it follows that the IIB KSE can be
written as
\bea
\label{killmax1}
\sum_{j=1}^{2m}(f^{-1}\partial_M f)_{ij} \eta_j+ {\cal D}_M \eta_i=0~,
\cr
P_A \Gamma^A \eta_i + \tfrac{1}{24} G_{ABC} \Gamma^{ABC} \eta_i=0~,
\eea
for $i=1, \dots, 2m$, where ${\cal D}_M$  is the
supercovariant derivative.
First consider the algebraic constraint given in ({\ref{killmax1}}).
Evaluating this constraint for $i=\ell$ and $i=\ell+m$ for $\ell=1, \dots, m$
we obtain
\bea
\label{killfact1}
P_A \Gamma^A \eta_\ell &=& 0~,
\cr
G_{ABC} \Gamma^{ABC} \eta_\ell &=&0 \ .
\eea

Next, by using the constraint on $G$ given in ({\ref{killfact1}}),
note that the supercovariant derivative acting on $\eta_i$, $i=1, \dots , 2m$,
simplifies to give
\bea
{\cal D}_M \eta_\ell &=& \nabla_M \eta_\ell
+\tfrac{i}{48} F_{MN_1 N_2 N_3 N_4} \Gamma^{N_1 N_2 N_3 N_4} \eta_\ell
+\tfrac{1}{8} G_{MAB} \Gamma^{AB} \eta_\ell~,\nonumber\\
{\cal D}_M \eta_{\ell+m} &=& i \nabla_M \eta_\ell
-\tfrac{1}{48} F_{MN_1 N_2 N_3 N_4} \Gamma^{N_1 N_2 N_3 N_4} \eta_\ell
-\tfrac{i}{8} G_{MAB} \Gamma^{AB} \eta_\ell~,
\eea
for $\ell=1, \dots, m$, where $\nabla_M = \partial_M -\frac{i}{2}Q_M + {1 \over 4} \Omega_{M,AB} \Gamma^{AB}$.
Substituting these expressions back into the gravitino Killing spinor equation in
({\ref{killmax1}}) and evaluating for $i=\ell$ and $i=\ell+m$ we find the
conditions
\bea
\label{gravfact}
\tfrac{1}{2}[ \sum_{j=1}^{2m}(f^{-1}\partial_M f)_{\ell j}~ \eta_j-i\sum_{j=1}^{2m}(f^{-1}\partial_M f)_{(\ell+m)j}~\eta_j]\qquad\qquad\qquad\qquad&&\nonumber\\
\qquad\qquad\qquad\qquad +\nabla_M \eta_\ell
+\tfrac{i}{48} \Gamma^{N_1\dots N_4 }
 F_{N_1\dots N_4 M} \eta_\ell&=&0\,,\nonumber \\
\sum_{j=1}^{2m}(f^{-1}\partial_M f)_{\ell j}~ \eta_j+i\sum_{j=1}^{2m}(f^{-1}\partial _M f)_{(\ell+m)j}~\eta_j+\tfrac{1}{4} G_{MBC} \Gamma^{BC}\eta_\ell&=&0 \ .
\eea

Hence we observe that the KSE
can be split into equations involving only $P$, or $G$, or $\Omega$ and $F$.
The fact that the equations factorize
in this fashion means that the geometry and fluxes of the solutions
are significantly constrained. It has been shown that the integrability
conditions of the KSE of maximally supersymmetric $G$-backgrounds
also factorize \cite{systiib}.

On evaluating the constraints ({\ref{gravfact}}) together with
({\ref{killfact1}}), one obtains a parallel transport equation
for the matrix $f$ of the form
\be
\label{parallel}
(f^{-1} \partial_M f)_{ij} + (C_M)_{ij}=0~,
\ee
where $C$ is a connection whose components are obtained from the spacetime Levi-Civita
connection and the fluxes of the supergravity theory.
This connection is the restriction of the supergravity supercovariant connection
to the subbundle of the Killing spinors. A necessary condition for the
condition ({\ref{parallel}}) to admit a solution is that the curvature $F(C)$
associated with $C$ vanish
\be
F(C) \equiv dC - C \wedge C =0 \ .
\ee
In general, it is not possible to choose $f$ to be the identity matrix, although it is possible
to pre-multiply $f$ by an arbitrary invertible constant matrix, as ({\ref{parallel}})
is invariant under the transformation $f \rightarrow gf$, where $g$ is an invertible constant
$2m \times 2m$ matrix.

It has been shown in \cite{giib} that the form of the maximally supersymmetric
$G$-background solutions depends on whether $G$ is compact or non-compact.
In the non-compact cases with $G= K \ltimes \bR^8$ for $K=Spin(7), SU(4), Sp(2),
SU(2) \times SU(2)$ or $K=\{ 1 \}$, the spacetime geometry always admits
a null vector field $X=e^-$ which is parallel with respect to the Levi-Civita
connection, $\nabla X=0$; and the holonomy of the Levi-Civita connection is contained in $K\ltimes\bR^8$,
${\rm hol}(\nabla)\subseteq K \ltimes \bR^8$. Co-ordinates $u,v, y^I$ for $I=1, \dots , 8$ can be chosen such that
$X={\partial \over \partial u}$, and the spacetime metric can be written as
\be
ds^2 =2 dv (du + V dv + n_I dy^I) + \gamma_{IJ} dy^I dy^J~,
\ee
where $V=V(v,y), n_I = n_I(v,y), \gamma_{IJ}=\gamma_{IJ} (v,y)$, \ie the spacetime is a pp-wave.
The above holonomy condition implies that the holonomy of the Levi-Civita connection
of the transverse 8-manifold, $Y_8$, defined by $u,v={\rm const}$, is contained in $K$.
In addition, the fluxes take the form
\be
P = P_- e^- , \quad G = e^- \wedge L, \quad F = e^- \wedge M~,~~~e^-=dv\,,
\ee
where $L$ is a 2-form on $Y^8$, and $M$ is a self-dual 4-form on $Y^8$ and may depend on both $y^I,v$.

For example, the fluxes for solutions with $G= Spin(7) \ltimes \bR^8$ are
\be
G = e^- \wedge L^{\mathfrak{spin}(7)}, \qquad F = e^- \wedge ({1 \over 14} Q_- \psi + M^{\bf{27}})~,
\ee
where $L^{\mathfrak{spin}(7)} \in \mathfrak{spin}(7)$, $M^{\bf{27}}$ lies in the {\bf{27}} irreducible representation
in the decomposition of 4-forms with respect to $Spin(7)$ and $\psi$ is the $Spin(7)$-invariant four-form.

The fluxes for the case $G=SU(4) \ltimes \bR^8$ are given by
\bea
G &=& e^- \wedge (L^{\mathfrak{su}(4)} + \ell \omega)~,
\cr
F &=& e^- \wedge (-{1 \over 12} Q_- \omega \wedge \omega + {\rm Re} (m \chi) + {\tilde{M}})~,
\eea
where $L^{\mathfrak{su}(4)} \in \mathfrak{su}(4)$, $\omega$ is the Hermitian (1,1) form, $\chi$ is the $SU(4)$-invariant (4,0) form
and ${\tilde{M}}$ is a traceless (2,2) form.
The remaining fluxes for non-compact $G$ are presented in \cite{giib}.

In addition a straightforward consequence of the results\footnote{The recently announced solution in \cite{sabra} is included.}
of \cite{giib} is that the most general solution in the $\bR^8$ case,
assuming that the transverse metric does not depend on $v$,
is\footnote{We can without loss of generality set $n=0$ in the corresponding solution of \cite{giib}.}
\bea
&&ds^2 =2 dv (du + V dv ) + \delta_{IJ} dy^I dy^J~,
\cr
&&P = P_-(v) e^- , \quad G = e^- \wedge L, \quad F = e^- \wedge M~,~~~
\eea
where
\bea
&&L={1\over2}L_{IJ}(v) dy^I\wedge dy^J~,~~~
M={1\over4!} M_{I_1\dots I_4}(v) dy^{I_1}\wedge\dots\wedge dy^{I_4}\,,
\cr
&&V={1\over2}A_{IJ}(v) y^I y^J+B_I(v) y^I+ H(y, v)~,~~~\partial_I^2 H(y,v)=0~,~~~
\cr
&&{\rm tr} A=-{1\over 6} M_{I_1\dots I_4} M^{I_1\dots I_4}-{1\over4} L_{IJ} {}^*L^{IJ}-2P_- P_-^*\,,
\eea
\ie $A_{IJ}$, $B_I$, $P$, $G$ and $F$ depend only on $v$ but otherwise unrestricted,  $M$ is self-dual in $\bR^8$ and
 $H$ is a harmonic function in $\bR^8$, \eg $H=b(v)+\sum_k{a_k(v)\over |y-y_k(v)|^6}$. This is the most general pp-wave
 solution of IIB supergravity, under the assumption mentioned above, which preserves at least 16 supersymmetries.

Next let us turn to backgrounds with Killing spinors invariant under compact $G$ groups.
 There are several cases for each choice of group $G$. All  such backgrounds can be found in
 \cite{giib}. The simplest case is for $G=G_2$ where
\be
ds^2 = ds^2(\bR^{2,1}) + ds^2(Y^7)~,~~~G=P=F=0~,
\ee
where $Y^7$ is a $G_2$-holonomy manifold.

As another example consider $G=SU(3)$. The spacetime geometry is the product of a four-dimensional
Lorentzian symmetric space with a Calabi-Yau manifold $Y^6$.
The supersymmetry conditions on the fluxes imply that
$P=G=0$. There are then three sub-cases to consider. First
$M=AdS_2\times S^2\times Y^6$, and the metric and fluxes are
\bea
&&ds^2=ds^2(AdS_2)+ds^2(S^2)+ ds^2(Y^6)~, \cr
&&ds^2(AdS_2)=-(e^0)^2+(e^1)^2~,~~~ds^2(S^2)=(e^5)^2+(e^6)^2~, \cr
&&F= {1 \over 2\sqrt{2}}[H^1 \wedge {\rm Re}\chi- H^2\wedge {\rm Im}
\chi]~,~~~\chi=(e^2+ie^7)\wedge (e^3+ie^8)\wedge (e^4+ie^9)~, \cr
&&H^1=\lambda_1\, e^0\wedge e^1+\lambda_2\, e^5\wedge
e^6~,~~~H^2=-\lambda_1\, e^5\wedge e^6+\lambda_2\, e^0\wedge e^1~,
\eea
for constants $\lambda_1, \lambda_2$, and the scalar curvature
of $AdS_2$ and $S^2$
are $R_{AdS_2}=-R_{S^2}=-4(\lambda_1^2+\lambda_2^2)$.

In the second case $M=CW_4(-2\m^2 {\bf 1})\times Y^6$, and the metric and fluxes are
\bea &&ds^2=ds^2(CW_4)+ ds^2(Y^6)~, \cr &&F={1 \over
2\sqrt{2}}[H^1 \wedge {\rm Re}\chi- H^2\wedge {\rm Im} \chi]~, \cr
&&H^1=\m\, e^-\wedge e^1~,~~~H^2=\m\, e^-\wedge e^6~, \eea
where $CW_4$ is the four-dimensional Cahen-Wallach space.

In the third case, $M=\bR^{3,1}\times Y^6$, and the metric and fluxes are
\bea
&&ds^2(M)=ds^2(\bR^{3,1})+ ds^2(Y^6)~,
\cr
&&F=0~.
\eea

Further details concerning these solutions as well as the solutions for $G=SU(2)$ can be found in \cite{giib}.
We remark that a key part of the computation is the understanding of the flatness condition of the $C$ connection. It is not always possible
to set $f=1$. This  can only happen whenever $C$ takes values in a subalgebra of $Spin(9,1)$ which preserves $\Delta^G$.
This is the case for $G=G_2$.

\newsection{Near maximal supersymmetry}

The method of spinorial geometry has recently been adapted to backgrounds with near maximal number of supersymmetries \cite{iibpreons,11preons}. The basic idea is that instead of specifying $N$ Killing spinors we can specify $N_{max}-N$ ``normal'' spinors, where $N_{max}$ is the maximal number of supersymmetries for the theory under study, \eg{}$N_{max}=32$ for IIB and $D=11$ supergravity. The gauge symmetry can then be used to simplify the form of the normal spinor(s), which in turn also simplifies the expressions for the Killing spinors. For $N=31$ we need to study 3 cases, corresponding to the number of orbits of the normal spinors under the action of the gauge group, in type IIB and 2 cases in $D=11$ supergravity. Essential for the definition of normal spinors is the existence of a non-degenerate pairing between spinors and normal spinors, see \cite{iibpreons} for details.


\subsection{$N=31$ in IIB supergravity}

This case simplifies as it turns out to be enough to study the algebraic KSE as we will see below. The Killing spinors can be written as
\be
      \e_r=\sum_{i=1}^{32} f^i{}_r \eta_i~,~~~r=1,\dots, N \,,\qquad
      \la{iibkilling}
\ee
where $\eta_p$, $p=1,\dots,16$, is a basis in the space of positive chirality
Majorana-Weyl spinors, $\eta_{16+p}= i\eta_p$ and $f$ are real spacetime functions.
Consider first a normal spinor in the orbit with stability subgroup $Spin(7)\ltimes\bR^8$. A representative for this orbit is
\be
       \nu=(n+im)(e_5+e_{12345})~.
\ee
Rewriting (\ref{iibkilling}) as
\be
       \epsilon_r= f^1{}_r (1+e_{1234})+ f^{17}{}_r i (1+e_{1234})+ f^k{}_r \eta_k \,,
\ee
where $\eta_k$ are the remaining basis elements,
and using the orthogonality condition between $\nu$ and $\epsilon_r$, we get
\be
      \epsilon_r=\frac{f^{17}{}_r}{n} (m+in) (1+e_{1234})+ f^k{}_r \eta_k~.
\ee
Note that the orthogonality relation has allowed us to eliminate one function from each Killing spinor.
    Substituting the form of $\e_r$ into the algebraic KSE and using that the rank of
    the matrix $(f^i{}_r)$ is 31, one finds
    \bea
      &&\hspace*{-0.4cm}P_M\Gamma^M C*[(m+in) (1+e_{1234})]+\tfrac{1}{24} G_{M_1M_2M_3}\Gamma^{M_1M_2M_3} (m+in) (1+e_{1234})=0~,\nonumber\\
      \vspace*{0.3cm}
      &&\hspace*{-0.4cm}P_M\Gamma^M\eta_p=0~,~~~~~~~G_{M_1M_2M_3}\Gamma^{M_1M_2M_3}\eta_p=0~,~~~\qquad p=2,\dots, 16~.
    \eea
   The factorization of the constraints on $P$ and $G$ is similar to that occurring for the maximal $G$-backgrounds
   as explained in the previous section.
    Noting that $\eta_p$ is {\it only} annihilated by either $\Gamma^+$ or $\Gamma^-$ implies that the
    only non-vanishing component of $P$ is either $P_+$ or $P_-$, respectively. Since both types of spinors occur $P=0$.
    Using that $P=0$, we find that $G_{M_1M_2M_3}\Gamma^{M_1M_2M_3} \eta_i=0$ for all
    the basis spinors $\eta_i$, which implies that $G=0$.
    If both $P$ and $G$ vanish the gravitino KSE is linear over the complex numbers
    implying that there is always an even number of Killing spinors, hence $N=31$ implies $N=32$. The analysis
    for the other two orbits is analogous \cite{iibpreons}.
    To our knowledge this is the first example which demonstrates that there are restrictions on the number of
    supersymmetries of backgrounds in a maximal supergravity theory.
    A similar result was found afterwards for IIA supergravity \cite{Bandos:2006xz}.

\subsection{$N=31$ in $11$D supergravity}

In $D=11$ there is no algebraic KSE which makes the problem much harder to analyze.
Instead we have to solve the parallel transport equation
      \be
        {\cal D}_A\e_r=0~,~~~r=1,\dots,31~.
      \ee
It is convenient to study the integrability condition
      \be
        {\cal R}_{AB} \e_r=[{\cal D}_A, {\cal D}_B]\e_r=0~,
      \ee
since the constraints from satisfying the field equations and Bianchi identities can easily be incorporated. In particular
$\Gamma^N{\cal R}_{MN}$ is a linear combination of field equations and Bianchi identities and therefore necessarily vanishes.
If one can show that ${\cal R}_{AB}=0$, then the backgrounds will be (locally)  maximally supersymmetric \cite{jfgp}.

There are two ways of solving the integrability conditions.
The first is to expand the supercovariant curvature in (a basis of) gamma matrices
      \be
        {\cal R}_{MN,ab}=\sum^5_{k=1} \frac{1}{k!}(T_{MN}^k)_{A_1A_2\dots A_k}
        (\Gamma^{A_1A_2\dots A_k})_{ab}~,
      \ee
      and let the gamma matrices act on $\e_r$ and read off the conditions on $T$.
      The second, more economical, way is to make use of a spinorial basis and write the supercovariant curvature as
      \be
        {\cal R}_{MN,ab}= u_{MN}^r\,\eta_{r,a} \nu_b~,
      \ee
      where $\nu$ is the normal spinor, $\e_r= f^s{}_r \eta_s$ and $\eta_r$ is a basis
      in the space of Killing spinors.
     ${\cal R}_{MN}$ can thus be written in terms of 31 two-forms $u_{MN}^r$ which is consistent with the fact that the stability
     subgroup of 31 spinors in $SL(32,\bR)$ is $\bR^{31}$. $T$ and $u$ can easily be related by contracting the expressions above with gamma matrices.

  After solving the integrability conditions ${\cal R}_{MN}$ is expressed in terms of the 31 two-forms $u_{MN}^r$, which are
  further constrained by the field equations and Bianchi identities.
     In particular the requirement that   $\Gamma^N{\cal R}_{MN}$ has to vanish, as  linear combination of field equations
     and Bianchi identities,  implies
      \bea
      && (T^1_{MN})^N = 0~,~~~  (T^2_{MN})_P{}^N = 0~,~~~(T^1_{MP_1})_{P_2} + \tfrac{1}{2} (T^3_{MN})_{P_1 P_2}{}^N =  0~, \vspace*{0.1cm}\nonumber\\
      && (T^2_{M[P_1})_{P_2 P_3]} - \tfrac{1}{3} (T^4_{MN})_{P_1 P_2 P_3}{}^N =  0~,~~~
      (T^3_{M[P_1})_{P_2 P_3 P_4]} + \tfrac{1}{4} (T^5_{MN})_{P_1 \cdots P_4}{}^N =  0~, \vspace*{0.1cm}\nonumber\\
      && (T^4_{M[P_1})_{P_2 \cdots P_5]} - \tfrac{1}{5 \cdot 5!} \epsilon_{P_1 \cdots P_5}{}^{Q_1 \cdots Q_6}
      (T^5_{M Q_1})_{Q_2 \cdots Q_6}  =  0~.
      \eea
      From the explicit expressions for $T$ in terms of the physical fields it follows that
      \bea
        && (T^1_{MN})_P = (T^1_{[MN})_{P]} \,, \qquad
        (T^2_{MN})_{PQ} = (T^2_{PQ})_{MN} \,, \qquad
        (T^3_{[MN})_{PQR]} = 0 \,.\qquad
        \eea
      After showing that $T^1=0$, which implies that $F\wedge F=0$, we have that
      \be
      (T^3_{MN})_{PQR} = \tfrac{1}{6} (\nabla_M F_{NPQR} - \nabla_N F_{MPQR}) \,,
      \ee
      which finally implies that ${\cal R}_{AB}=0$ and thus maximal supersymmetry \cite{jfgp}.

\newsection{Concluding remarks}

In this review, we outlined some aspects of the spinorial geometry approach to solving the KSE
of supergravity theories. In addition, we presented some of the applications, like  the classification of maximally supersymmetric IIB
$G$-backgrounds  and $N=31$ supersymmetric IIB and $D=11$ supergravity backgrounds. We have also emphasized that we have constructed
the most general pp-wave solution of IIB supergravity which preserves at least 16 supersymmetries.

The question that remains is whether the supersymmetric backgrounds of IIB and $D=11$ supergravities can be classified\footnote{IIA backgrounds
are special cases of the $D=11$ ones.}.
There are two classes of supersymmetric backgrounds. One class consists of those backgrounds which admit   Killing spinors
that are invariant under some proper Lie subgroup of the appropriate $Spin$ gauge group. Examples of such backgrounds are those in IIB and $D=11$
supergravities  with
$N=1$ supersymmetry, and the
maximally supersymmetric IIB $G$-backgrounds that we have mentioned above. The other class is those backgrounds for which
the stability subgroup of their Killing spinors in the $Spin$ group is $\{1\}$. Examples of such backgrounds are those in IIB and
$D=11$ supergravities that preserve $N=31$ and $N=32$ supersymmetries. It can be shown that two spinors in IIB or $D=11$ supergravities
can have a trivial subgroup in the $Spin$ gauge groups. So there may exist such backgrounds for any $N\geq 2$. For the former class,
it seems likely that the KSE of backgrounds whose Killing spinors are invariant under a proper
Lie subgroup of $Spin$ can be solved. The invariance condition imposes a strong restriction on the form of Killing spinors
and so the linear systems that arise from the spinorial geometry can be tractable. For the latter class, it is  encouraging
that the $N=31$ backgrounds we have investigated are maximally supersymmetric. This indicates that if the Killing spinors are not invariant
under some proper Lie subgroup of $Spin$, then the KSE together with the field equations and the Bianchi identities
 impose strong restrictions on the geometry and fluxes
of the background. If  this persists  for backgrounds with fewer supersymmetries, then
 the classification of supersymmetric backgrounds is simplified and so the programme can be carried out in full.

\section*{Acknowledgments}

Part of the work this review is based on was completed while D.R.~was a post-doc at King's College
London, for which he would like to acknowledge the PPARC grant
PPA/G/O/2002/00475. In addition, he is presently supported by the European
EC-RTN project MRTN-CT-2004-005104, MCYT FPA 2004-04582-C02-01 and CIRIT GC
2005SGR-00564. U.G.~has a
postdoctoral fellowship funded by the Research Foundation
K.U.~Leuven.

\end{document}